\documentclass[12pt]{iopart}

\usepackage{graphicx}
\usepackage{dcolumn}
\usepackage{bm}
\usepackage{color}
\usepackage{ulem}
\begin{document}
\title{Strong pinning of vortices by antiferromagnetic domain boundaries in CeCo(In$_{1-x}$Cd$_x$)$_5$}
\author{Dong-Jin Jang$^{1}$\footnotemark[3]
, Luis Pedrero$^1$, L. D. Pham$^2$, Z. Fisk$^3$ and Manuel Brando$^{1}$\footnotemark[3]\footnotetext{Corresponding authors\\Email: dongjin.jang@cpfs.mpg.de, manuel.brando@cpfs.mpg.de}}
\address{$^1$ Max Planck Institut f{\"u}r Chemische Physik fester Stoffe, 01187 Dresden, Germany}
\address{$^2$ University of California, Davis, California 95616, USA}
\address{$^3$ University of California, Irvine, California 92697-4573, USA}  

\date{\today}

\begin{abstract}
We have studied the isothermal magnetization $M(H)$ of CeCo(In$_{1-x}$Cd$_x$)$_5$ with $x$ = 0.0075 and 0.01 down to 50\,mK. Pronounced field-history dependent phenomena occur in the coexistence regime of the superconducting and antiferromagnetic phases. At low-fields, a phenomenological model of magnetic-flux entry well explains $M(H)$ implying the dominance of bulk pinning effect. However, unless crystallographic quenched disorder is hysteretic, the asymmetric peak effect (ASPE) which appears at higher fields cannot be explained by the pinning of vortices due to material defects. Also the temperature dependence of the ASPE deviates from the conventional scenario for the peak effect. Comparison of our thermodynamic phase diagrams with those from previous neutron scattering and magnetoresistance experiments indicates that the pinning of vortices takes place at the field-history dependent antiferromagnetic domain boundaries.
\end{abstract}
\maketitle 

\section{Introduction}
CeCoIn$_5$ is the highest $T_\textrm{c}$ superconductor among the Ce-based heavy-fermion compounds. It shows a unique correlation between superconducting (SC) and antiferromagnetic (AF) phases at the verge of a quantum critical point (QCP)~\cite{Petrovic,Bauer}. The existence of a QCP with notable AF fluctuations upon escaping SC ground state~\cite{Paglione,Bianchi}, the unconventional order parameter for the SC phase~\cite{Radovan,Movshovich,Kenzelmann,Young,Izawa}, and instabilities of both SC and AF phases revealed by substitution studies~\cite{Bauer,Daniel,Pham,Nair} have led to a general consensus that the SC phase is mediated by spin fluctuations and that the AF phase is strongly entangled with the SC phase. Specifically, a recent study on Cd-substituted CeCoIn$_5$ has provided plausible evidences that AF and SC phases might not only microscopically coexist in real space but also show phase separation in $\bm{k}$-space\cite{Nair}. 

Earlier studies show that Sn preferentially substitutes In in the CeIn(1)-plane and that the SC phase is largely suppressed even with a few percent of Sn~\cite{Daniel}. Similar to Sn-substitution, Cd occupies In(1) and In(2) sites~\cite{Booth} and causes a substantial suppression of the SC phase~\cite{Pham}. Unlike Sn-substitution, however, Cd introduces droplets of antiferromagnetism in a SC matrix~\cite{Urbano} and a long-range AF phase develops above the SC phase boundary as the substitution level becomes higher than about 0.6~\%~\cite{Nair}. The QCP and the 1$^\textrm{st}$-order SC phase transition due to the Pauli-limiting are smeared out upon Cd substitution~\cite{Pham,Nair}. 

Among the large number of studies on CeCoIn$_5$ and its isostructural relatives CeRhIn$_5$ and CeIrIn$_5$, a relatively small number of studies have focused on the dynamics of vortex matter observed in magnetization $M(H)$, essentially for two reasons: first, the 1$^\textrm{st}$-order SC phase transition at high fields in conjunction with an exotic high-field and low-temperature phase ($\mathbf{Q}$-phase) has been the main focus of scientific attention~\cite{Radovan,Kenzelmann,Bianchi1}; second, the isothermal magnetization shows a peculiar and complex behavior which is still not fully understood~\cite{Tayama,Murphy,Gratens}. 
 
Here, we report isothermal $M(H)$ measurements on CeCo(In$_{1-x}$Cd$_x$)$_5$ with $x$ = 0.0075 and 0.01. The measurements were performed down to 50\,mK in an external magnetic field $\mu_0H$ (up to 12\,T) parallel and perpendicular to the crystallographic $\mathbf{c}$-axis. Features which originate from the SC order are very pronounced in $M(H)$, while signatures of the AF phase are weaker, but they have clearly been observed in other techniques like nuclear magnetic resonance~\cite{Urbano}, neutron scattering~\cite{Nair} and magnetoresistance~\cite{Nair}. We observe in $M(H)$ an unusual asymmetric peak effect (ASPE) and typical hysteresis of a type-II superconductor~\cite{Tayama,Murphy,Gratens}. The classical analysis of $M(H)$ with a phenomenological model of flux-entry reveals that the bulk pinning of vortices causes strong hysteresis at low fields. On the other hand, the crystallographic quenched disorder induced by Cd substitution cannot cause the observed ASPE, because in this case the pinning energy would have no field-dependent hysteresis. From our $M(H)$ measurements, we derive the $T - H$ phase diagrams for $\mathbf{H \parallel c}$ and $\mathbf{H \perp c}$ and compare them with those found from other experimental techniques. We see that the temperature dependence of the ASPE mimics the temperature dependence of the AF domain transition and that the weakening of the ASPE coincides with the attenuation of the signature attributed to the hysteresis due to AF domains. This implies that the field-history dependent AF domain boundaries are responsible for the strong bulk pinning of vortices. 

\begin{figure*}
\includegraphics[width=\textwidth]{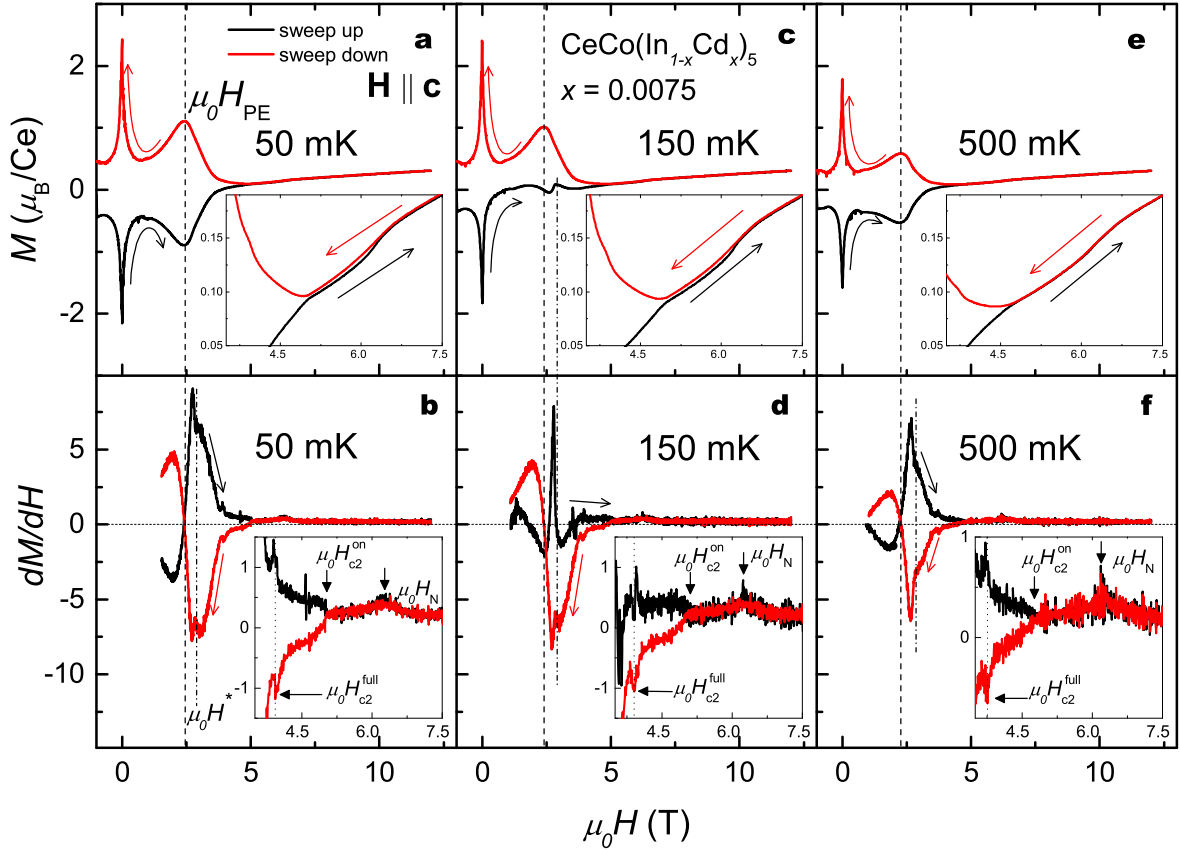}
\caption{(color online) {\bf Isothermal magnetization of CeCo(In$_{0.9925}$Cd$_{0.0075}$)$_5$ with $\mathbf{H \parallel c}$.} Panel {\bf a} shows $M(H)$ at 50\,mK. Black and red solid lines are for up and down sweeps, respectively. The arrows alongside the experimental data also identify the sweep directions. The inset is a magnified view in the field range 3.5-7.5\,T. Panel {\bf b} shows differential susceptibility $dM/dH$ which is numerically derived from panel {\bf a}. The vertical dashed-line crossing the two main panels locates the exact position of the extremes of the PE at $\mu_0H_\textrm{PE}$. The vertical dash-dotted line with arrow in panel {\bf b} is placed at $\mu_0 H^\textrm{*}$ where the local maximum of $dM/dH$ in decreasing field is found near $\mu_0H_\textrm{PE}$. The inset of panel {\bf b} shows further magnification of $dM/dH$ where signatures of critical fields for the AF phase ($\mu_0 H_\textrm{N}$), onset and full development of the SC phase ($\mu_0H_\textrm{c2}^\textrm{on}$ and $\mu_0H_\textrm{c2}^\textrm{full}$, respectively) are accentuated. Panels {\bf c} and {\bf d} show results at 150\,mK. Panels {\bf e} and {\bf f} are results at 500\,mK. At 150\,mK, the PE is absent in increasing field and the anomaly at $\mu_0 H^\textrm{*}$ becomes pronounced. The extended dash-dotted line from {\bf d} to {\bf c} notes that the local maximum of the ripple in $M(H)$ is at $\mu_0 H^\textrm{*}$.}
\label{fig1}
\end{figure*}

\section{Experiments}
The single crystals used in our experiments were prepared by standard In-flux technique as described in Ref. 11. The microprobe analysis confirmed that 10 times of stoichiometric amount of Cd must be used to get the composition of CeCo(In$_{1-x}$Cd$_x$)$_5$. In this work, we investigated two compositions, $x$ = 0.0075 and $x$ = 0.01. For the measurements, 10 mg of plate-like samples were mounted in the Faraday-force magnetometer as described in Ref. 19. This apparatus consists of one pair of metallic plates (a capacitor) normal to a uniform magnetic field and a gradient field. We used here a gradient field of 10 T/m. The force on the sample is $\mathbf{F} =- (\mathbf{M} \cdot \nabla) \mathbf{B}$ and this is proportional to the change in the measured capacitance. The magnetic units have been obtained by comparing the data with data obtained by using SQUID.
\section{Results} 
CeCoIn$_5$ becomes a superconductor below $T_\textrm{c}$=2.3\,K~\cite{Petrovic}. Cd substitution not only weakens the SC phase but also induces an AF phase. In CeCo(Cd$_{x}$In$_{1-x}$)$_{5}$, the N\'{e}el temperature, $T_\textrm{N}$, becomes larger than $T_\textrm{c}$ for $x\geq$ 0.006. At $x$ = 0.0075, $T_\textrm{N}$ is 2.4\,K and $T_\textrm{c}$ is 1.7\,K~\cite{Nair}. The critical field for the AF phase is correspondingly larger than that of the SC phase and both are anisotropic as in pure CeCoIn$_{5}$.

Fig.~\ref{fig1} shows representative $M(H)$ curves for CeCo(In$_{0.9925}$Cd$_{0.0075}$)$_5$ with $\mathbf{H \parallel c}$. At 50 mK we observe a tiny kink at the AF transition field $\mu_0H_\textrm{N} = 6.3$\,T (see inset of Fig.~\ref{fig1}a) and a large hysteresis loop which includes sharp peaks near zero field and a fish-tail shape in the intermediate field range. This hysteresis is obviously resulting from vortices. Especially, the fish-tail in $M(H)$ is a well known hallmark of the peak effect. The peak effect is generally attributed to an enhanced bulk pinning over a softened vortex-vortex interaction~\cite{Larkin,Blatter}. Once this effect is initiated, increasing (decreasing) external field cannot easily force vortices to enter (to escape from) a superconductor and the magnetization is decreased (increased) as ${\bf M} =\frac{1}{\mu_0}{\bf B}-{\bf H}$. However, its manifestation in various kinds of superconductors requires a heuristic approach to understand what is triggering the enhancement of the bulk pinning strength. 

\begin{figure*}
\includegraphics[width=\textwidth]{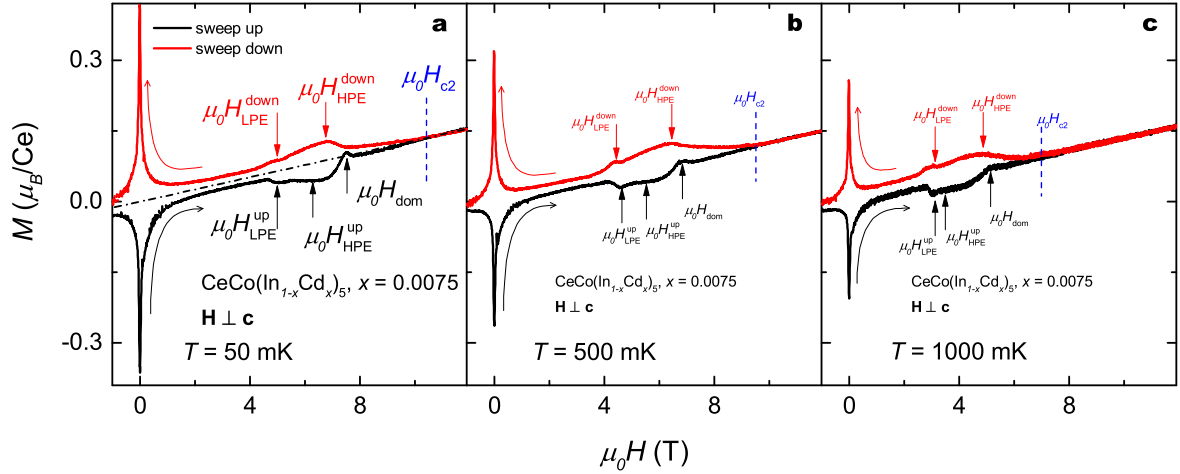}
\caption{(color online) {\bf Isothermal magnetization of CeCeCo(In$_{0.9925}$Cd$_0.0075$)$_5$ with $\mathbf{H \perp c}$.} Panels {\bf a}-{\bf c} exhibit $M(H)$ curves at temperatures of 50, 500, and 1000\,mK, respectively. Red and black solid lines indicate field sweeps up and down, respectively. Peaks are visible between 4 and 8\,T on top of the peak effect. We refer to those peaks as ${\mu_0H}_\textrm{LPE}^\textrm{up}$ (${\mu_0H}_\textrm{HPE}^\textrm{up}$) and ${\mu_0H}_\textrm{LPE}^\textrm{down}$ (${\mu_0H}_\textrm{HPE}^\textrm{down}$): these field designate the lower (higher) peak positions of the double peaks on sweeping up and down, respectively. The upper critical field, $\mu_0 H_\textrm{c2}$ (blue dashed line), is located at the end of the hysteresis and the value matches with the one found in Ref. 12. The first local maximum below $\mu_0 H_\textrm{c2}$ in the sweep-up data is labeled $\mu_0H_\textrm{dom}$ because we believe that the rather abrupt change in field-history dependent AF domain population causes this positive peak (see the discussion). The linear dash-dot line in Fig.~\ref{fig2}a is the paramagnetic background which is subtracted to obtain the data in Fig.~\ref{fig3}a.}
\label{fig2}
\end{figure*}

Beside the peak effect which is characterized by the extrema indicated by $H_\textrm{PE}$, there are several other features in $M(H)$ that deserve attention. To determine the field values of those features we show the magnetic susceptibility $dM/dH$ in Fig.~\ref{fig1}b. The kink in $dM/dH$ at $\mu_0H_\textrm{N}$ identifies the antiferromagnetic-to-paramagnetic transition as deduced from neutron scattering and magnetoresistivity experiments~\cite{Nair}. In particular, Nair \textit{et al.}~\cite{Nair} have observed a monotonic decrease of the resistivity below $\mu_0H_\textrm{N}$ followed by a small peak; as the field is further reduced, the resistivity drops sharply indicating the transition into the superconducting state. We refer to this field as $\mu_0H_\textrm{c2}^\textrm{on}$. At an even smaller field the resistivity drops to zero and we refer to this field as $\mu_0H_\textrm{c2}^\textrm{full}$. We observe the signatures at these fields in $dM/dH$ (lower panels of Fig.~\ref{fig1}). At $\mu_0H_\textrm{c2}^\textrm{on}$, the hysteresis loop opens, and clear spikes are observed in $dM/dH$ at $\mu_0H_\textrm{c2}^\textrm{full}$, see the insets of Fig.~\ref{fig1}b, d, and f. We also observe another feature in $dM/dH$ at a slightly higher field than $\mu_0H_\textrm{PE}$ which we label $\mu_0H^\textrm{*}$. Since this feature is very weak and it was not seen in other experiments, it is difficult to draw conclusions about its nature. While most of the features in $M(H)$ at 50 mK survive at elevated temperatures, the peak effect substantially weakens and the anomaly at $\mu_0H^\textrm{*}$ becomes more pronounced (see Fig.~\ref{fig1}c and d). We also found that between 100 and 300\,mK the peak effect appears to be almost absent when $H$ is increased but strong when $H$ is decreased. The ASPE and the asymmetric anomaly at $\mu_0H^\textrm{*}$ both become symmetric at elevated temperatures above 500 mK (see Fig.~\ref{fig1}e and f).

\begin{figure*}
\includegraphics[width=\textwidth]{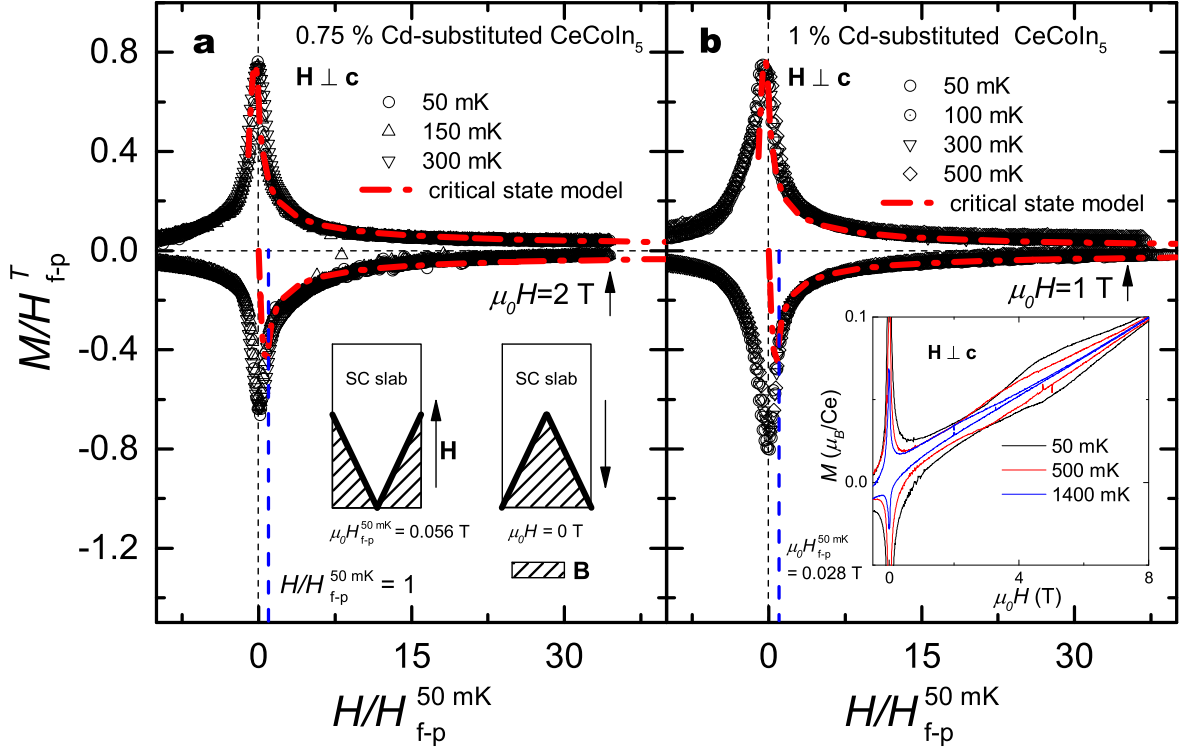}
\caption{(color online) {\bf Critical state model applied at low fields.} {\bf a} Low-$T$ $M(H)$ curve after having subtracted a linear paramagnetic contribution for $\mathbf {H \perp c}$. The curves are scaled by using the critical state model of vortex entry (red dash-dotted line). The best fit to the model returns the full penetration field, $\mu_0H_\textrm{f-p}^{T}$ = 0.056\,T at $T$ = 50\,mK. At this field, the initial `V'-shaped field profile is configured as in the left picture of the inset: the hatched area describes the internal field distribution inside a SC slab. Due to the bulk pinning of vortices, substantial amount of flux remains even if the field is swept back to zero as depicted in the right inset of panel {\bf a}. {\bf b} The same analysis as in panel {\bf a} for the sample with $x = 0.01$. The inset shows $M(H)$ curves with 1~\% Cd. It should be stressed that all the asymmetric features shown in Fig.~\ref{fig2} are gone by adding only 0.25~\% more Cd. }
\label{fig3}
\end{figure*}

Figs.~\ref{fig2}a-c exhibit $M(H)$ curves with $\mathbf{H \perp c}$ at temperatures of 50, 500, and 1000\,mK. Because of both magnetocrystalline and SC anisotropy, the maximum value of the magnetization only reaches 0.42 $\mu_B$/Ce near zero field (Fig.~\ref{fig2}a) which is about a factor of 6 smaller than the value found for $\mathbf{H \parallel c}$ (Fig.~\ref{fig1}a).
We used the same method of comparison between magnetoresistance and $M(H)$ for $\mathbf{H \perp c}$ as we did for with $\mathbf{H \parallel c}$ to be able to identify the characteristic fields. In $M(H)$, the hysteresis closes above $\mu_0H_\textrm{c2}$ (Fig.~\ref{fig2}). It is worth noting that the field where the two $M(H)$ curves for opposite field sweep directions merge is technically defined as irreversibility field. However, here the upper critical field practically coincides with the irreversibility field. For $\mathbf{H \perp c}$ the peak effect is asymmetric in a wide range of the temperatures. Moreover, there are several peaks in either field sweep directions: $\mu_0H_\textrm{LPE}^\textrm{up}$ ($\mu_0H_\textrm{HPE}^\textrm{up}$) and $\mu_0H_\textrm{LPE}^\textrm{down}$  ($\mu_0H_\textrm{HPE}^\textrm{down}$) indicate the peak positions of the lower (higher) peak when the external field is increased and decreased, respectively. The correct positions of them are determined by the numerical derivatives as it was done in Fig.~\ref{fig1}. Importantly, a single positive peak develops only in increasing field and it will be discussed later in connection with the hysteretic AF domain structure. We label this positive peak $\mu_0H_\textrm{dom}$.

\begin{figure*}
\includegraphics[width=\textwidth]{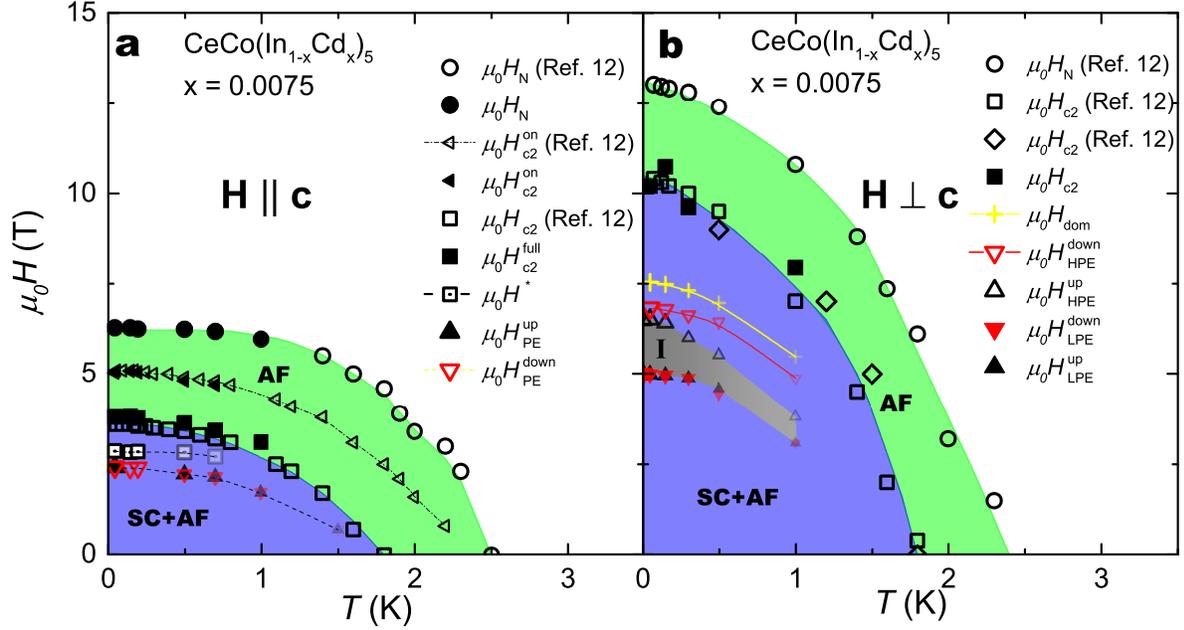}
\caption{ (color online) {\bf $H$-$T$ phase diagrams of CeCo(In$_{0.9925}$Cd$_{0.0075}$)$_5$.} Panel {\bf a} is the phase diagram with $\mathbf{H \parallel c}$. In the green region the AF phase dominates and in the blue region the AF and SC phases coexist. The phase boundaries are guide lines based on $M(H)$ and other points reproduced from Ref. 12. $\mu_0 H_\textrm{c2}^\textrm{on}$ (marked by empty left triangles in the middle of the AF phase) is the point where a sharp drop of magnetoresistance appears and it coincides well with $\mu_0 H_\textrm{c2}^\textrm{on}$ found Fig.~\ref{fig1} (filled left triangles). Zero resistance marks the full development of the superconductivity (empty square), and it is consistent with $\mu_0 H_\textrm{c2}^\textrm{full}$ (filled square) defined in Fig.~\ref{fig1}. $\mu_0 H^\textrm{*}$ (dotted-square) is a new characteristic field which is only found in $M(H)$. The upper (lower) triangles are the peak positions of the peak effect in increasing (decreasing) field. Symbols which fade away reflect the observation that features are weakened in increasing $T$. Panel {\bf b} is the phase diagram with $\mathbf{H \perp c}$. $\mu_0 H_\textrm{dom}$ (yellow cross mark) designates a small positive peak in $M(H)$ observed only in increasing field as shown in Fig.~\ref{fig2}. The grey area labeled as I is located within the two peak effect fields which tend to merge above 1\,K only in increasing field.}
\label{fig4}
\end{figure*}

We take now a closer look at the low-field region where a sharp hysteresis loop is opened. This major hysteresis loop at low fields can be well described by the critical state model of vortex entry into a superconducting slab~\cite{Bean,Teruo}. A detailed analysis will be shown for the case of $\mathbf{H \perp c}$ just because it satisfies the condition of the model which requires $\mathbf{H}$ to be along the long side of the plate-like sample. The observed hysteresis can be understood as follows: the bulk pinning force hinders a magnetic-flux to move in and out of the sample so that there remains an internal field trapped by vortices even though the external field is cycled back to zero. This is schematically described in the insets of Fig.~\ref{fig3}a. To apply the model in detail (see Appendix), we consider the data below $\mu_0 H_\textrm{LPE}^\textrm{up}$ and subtract a linear response (dash-dotted line in Fig.~\ref{fig2}a) which is irrelevant to the vortex dynamics. In this system, a linear paramagnetic contribution from the normal vortex core arises because the size of the vortex core, $\xi_\textrm{GL}$, is as large as 10\,nm in diameter~\cite{Nair} which encompasses many $4f$ moments. The most important parameter in this calculation is the full-penetration field $\mu_0H_\textrm{f-p}$ at which the internal field is finally reaches to the center of the slab and a `V'-shaped field distribution is formed (see the left inset of Fig.~\ref{fig3}a).

For $x$ = 0.0075, $\mu_0H_\textrm{f-p}^T$ (a full-penetration field at a certain temperature $T$) is 0.056\,T and the power $\gamma$ for the internal field dependent pinning force ($F_\textrm{p}\sim B^\gamma$) is about 0.4 at $T=$ 50\,mK. As the temperature is increased, $\mu_0H_\textrm{f-p}^T$ decreases monotonically but $\gamma$ remains more or less the same. Given that $\gamma$ decides the height of the hysteresis loop of a reduced magnetization $M/H_\textrm{f-p}^{T}$, and that $H_\textrm{f-p}^T$ decides the range of the input variable, $H/H_\textrm{f-p}^T$ for a fixed range of $H$, the maximum and the minimum of $M/H_\textrm{f-p}^{T}$ do not change much, but the field range of the model is widened as the temperature is increased. This observation allows us to determine a scaling behavior which is shown in Fig.~\ref{fig3}. In this figure scaled sets of data are generated by simply using a common factor, $H_{\textrm{f-p}}^{\textrm{50 mK}}$, for the reduced field. For the sample with $x = 0.01$ the calculation and the scaling apply equally well but $\mu_0H_\textrm{f-p}^\textrm{50 mK}$ is decreased to 0.028\,T. This implies that the Cd substitution indeed weakens the SC phase.
 
\section{Discussion} 
The magnetic phase diagrams for $x$ = 0.0075 are displayed in Fig.~\ref{fig4}: Fig.~\ref{fig4}a and b are for $\mathbf{H \parallel c}$ and $\mathbf{H \perp c}$, respectively. We included points from Ref. 12. In Fig.~\ref{fig4}a, the AF transition seen in $M(H)$ is marked by a filled-circle and the whole AF phase boundary is completed by points taken from Ref. 12 (circle). The onset of the superconductivity at $\mu_0H_\textrm{c2}^\textrm{on}$ is recognized by opening of a hysteresis loop in $M(H)$ from Fig.~\ref{fig1} (filled-triangle) and a vertical drop of the resistivity (triangle) in Ref. 12. The full development of the superconductivity is noted by zero resistance~\cite{Nair} (square) and simultaneously by a spike in $dM/dH$ from Fig.~\ref{fig1} (filled-square) at $\mu_0H_\textrm{c2}^\textrm{full}$. It is worth mentioning that the SC transition is second order and the Pauli-limiting behavior in pure compound is absent. Therefore, the first order SC transition of CeCoIn$_5$ in the low-temperature and high-field regime is very susceptible to the material defect. Nuclear-magnetic-resonance with $\mathbf {H \parallel c}$ indicated that staggered moments are induced around Cd-sites and interaction between those AF droplets eventually gives rise to long-range AF order as the Cd concentration is increased~\cite{Urbano}. From this result, spatially disordered nucleation of the SC order parameter is equally likely, and the large gap between $H^\textrm{on}_\textrm{c2}$ and $H^\textrm{full}_\textrm{c2}$ might be attributed to percolation of zero-resistance paths influenced by fluctuations of SC order parameter. 

Inside the SC+AF phase, magnetic anomalies appear at $\mu_0H^\textrm{*}$ (dotted squares shown in Fig.~\ref{fig4}a). Some resemblance of $\mu_0H^\textrm{*}(T)$ with the vortex phase diagram of the pure compound opens a possibility that $\mu_0H^\textrm{*}(T)$ might be associated with a structural transition of vortex lattice from hexagonal to rhombic one~\cite{Bianchi2}. Nevertheless, it should be underlined that the vortex lattice is distorted due to the strong bulk pinning. In addition, signatures for $\mu_0H^\textrm{*}$ depicted in Fig.~\ref{fig1} are not reversible at low temperatures. Also, the field and temperature dependencies of the structural transition of the vortex lattice in pure compound do not fit to those of $\mu_0H^\textrm{*}$. The weak feature at $\mu_0H^\textrm{*}$ fades away as the temperature is increased above 0.5\,K (as reflected in faint symbols in Fig.~\ref{fig4}a) and it seems to follow the similar $T$-$H$ dependence of the lines of the peak effect (upper and lower triangles in Fig.~\ref{fig4}a). Hence, it is reasonable to assume that $\mu_0H^\textrm{*}$ is closely related to the pinning mechanism. However, since this feature has never been observed in other experiments, we do not know its origin.

With $\mathbf{H \parallel c}$, the peak effect is very strong at low temperatures and it is weakened in increasing $T$ as reflected by faint symbols in Fig.~\ref{fig4}a. This is in contradiction to the well-known behaviors that peak effect typically appears in the vicinity of $H_\textrm{c2}$ and that it becomes stronger at higher temperatures if the pinning is weak but collective over vortex lattice~\cite{Larkin,Blatter,Mohan}. Furthermore, as we increase the Cd content to 1~\%, the the peak effect becomes symmetric and largely suppressed (see the inset of Fig.~\ref{fig3}b). Although the magnitude of the peak is not a monotonically increasing function of the density of pinning centers, it is worth to note that we are substituting the In-sites by Cd which selectively replace less than 1~\% of the In(1) and In(2) sites. Therefore, crystallographic disorder itself may not supply enough energy to deform the entire vortex lattice. Even though negligible substitution indeed introduces sizable random deformation of the vortex lattice by collective pinning, it is still apparent that static pinning centers without field-history dependence cannot contribute to the asymmetric features observed in $M(H)$. In consequence, the physically solid argument which we must rely on is that the pinning should be bulky and static at low fields as quantitatively described by the critical state model (Fig.~\ref{fig3}), but they are essentially dynamic in response to higher external fields. Concluding, crystallographic quenched disorder can definitely not be the origin of the ASPE.

As alternative, we propose that field-history dependent AF domain boundaries act as major pinning centers. Since AF and SC orders coexist, certain vortex structure is expected to exist in a single AF domain and vice versa. Then, it is reasonable to assume that substantial disorder is induced at AF domain boundaries because vortex structures with different orientations are overlapping at the boundaries and these are thick enough to contain many vortices. Previous results which show the existence of AF domain structures~\cite{Nair,Urbano} can be combined to conclude that the domain wall thickness is comparable to the vortex core size, $\xi_\textrm{GL}$. The assumption is further supported by inspecting the $T$-$H$ phase diagram with $\mathbf{H \perp c}$ in Fig.~\ref{fig4}b in conjunction with neutron scattering data. In fact, in Ref. 12, a hysteretic change in the AF domain population is reported: (1/2 1/2 1/2) Bragg intensity drops rapidly in increasing fields above $\mu_0H_\textrm{dom}$ and it increases slowly when the field is cycled back. This hysteresis provides an explanation for the positive peak at $\mu_0H_\textrm{dom}$ in Fig.~\ref{fig2} since the staggering of magnetic moments in a certain direction is rapidly released only in increasing field. More correspondences between features from neutron data and $M(H)$ can be found. For instance, at 0.5\,K, a dip develops in the (1/2 1/2 1/2) Bragg intensity around 5\,T with increasing field. In $M(H)$, the position of the dip is between $\mu_0H_\textrm{LPE}^\textrm{up}$ and $\mu_0H_\textrm{HPE}^\textrm{up}$ (Fig.~\ref{fig2}b). At 1.5\,K, the neutron intensity neither shows a dip nor a rapid drop along $H$ and all the anomalies in $M(H)$ vanish as implied by the faint symbols and the grey-colored region (it is labeled as I to distinguish up and down sweeps) in Fig.~\ref{fig4}b. In the phase diagram, the temperature dependence of the characteristic fields related to the ASPE mimic the same temperature dependence of $\mu_0H_\textrm{dom}$. This observation repeatedly corroborates our assumption from which we deduce that the pinning is dominated by AF domain structure. 

Our result seems not to agree with previous magnetization studies. In CeCoIn$_5$, the ASPE is strongest when $\mathbf{H \perp c}$~\cite{Tayama,Murphy,Gratens}. Anomalies related to the vortex dynamics observed in recent experiment are interpreted in terms of a surface effect which is assumed to be most effective when $\mathbf{H \perp c}$. In CeCo(In$_{1-x}$Cd$_x$)$_5$, however, the ASPE is most pronounced when $\mathbf{H \parallel c}$ (Fig.~\ref{fig1}) and the Cd substitution reverses the angle-dependent appearance of it. We also did not find experimental evidences that could support the surface effect in angle-dependent phenomena other than evidences supporting bulk anisotropic physical properties. It should be noted that the peak effect is in general a bulk phenomenon, and the surface effect we might expect is the edge contamination of the vortex lattice which screens a sharp order-disorder transition of the vortex matter~\cite{Paltiel,Paltiel1}.

Although neutron scattering is not yet available with $\mathbf{H \parallel c}$, we expect the ASPE in Fig.~\ref{fig1}b to originate also from hysteretic changes of AF domains so that vortices can be pinned either maximally or minimally depending on the average size and structure of the AF domain boundaries. The field-history dependent anomalies at $\mu_0H^\textrm{*}$ which has similar behavior to the line of the peak effect in the $T$-$H$ plane might also support our assertion (Fig.~\ref{fig4}a). Results presented here should motivate precise mapping of field and temperature dependencies of AF domain boundaries in Cd-substituted CeCoIn$_5$ to test the proposed pinning mechanism.

In summary, we have studied the low-temperature isothermal magnetization of Cd-substituted CeCo(In$_{1-x}$Cd$_{x}$)$_5$ with $x=$ 0.0075 and 0.01. Signatures of the antiferromagnetic state are observed, but they are weak. On the other hand, interplay between vortices and pinning centers are responsible for all major features seen in magnetization. The critical state model for the vortex entry was successfully applied to confirm that the bulk defect pinning prevails at low fields. At higher fields, pronounced asymmetric peak effect is developed. The field and temperature dependencies of the asymmetric peak effect deviate from the conventional scenario based on pinning by material defects but mimic those for the antiferromagnetic domain structure in the $T-H$ phase diagrams. In addition, the weakening of the asymmetric peak effect coincides with the attenuation of the signature associated with the existence of antiferromagnetic domains. These observations support the conclusion that strong pinning of vortices takes place at the field-history dependent antiferromagnetic domain boundaries.
\vspace{1cm} 
\section{Appendix: critical state model}

 The global balance between the magnetic pressure from the periphery of a superconductor and the bulk pinning force, $\mathbf{F_\textrm{p}}(B)$, as a function of an internal magnetic field is described below. 
\begin{equation}
\fl
\mathbf{\frac{1}{\mu_0}\left(\nabla \times B \right)\times B}={\mathbf {F_\textrm{p}}(B)}; \quad F_\textrm{p}(B)=\alpha B^\gamma,
\end{equation}

where $\alpha$ and $\gamma$ are constants. By assuming a simplified geometry of a superconducting slab, the full penetration field strength, $H_\textrm{f-p}$,

\begin{equation}
\fl
H_\textrm{f-p}=\frac{1}{\mu_0} \{\left(2-\gamma \right)\mu_0\alpha d \}^{1/(2-\gamma)},
\end{equation}

and the magnetization becomes,

\begin{equation}
\fl
M=\frac{1}{\mu_0 d} \int_0^d dxB(x)-H,
\end{equation}

where $2d$ is the thickness of the slab. Scaled unitless magnetization, $M/H_\textrm{f-p}$, as a function of the normalized external field strength, $h=H/H_\textrm{f-p}$, can be derived as follows,

\begin{equation}
\fl
\label{cases}
\frac{M}{H_\textrm{f-p}}=\cases{\frac{2-\gamma}{3-\gamma} h^{3-\gamma}-h &  $,H_\textrm{f-p} \leq H,$ \\
\frac{2-\gamma}{3-\gamma} \{h^{3-\gamma}-(h^{2-\gamma}-1)^{\frac{3-\gamma}{2-\gamma}} \}-h  & $,H_\textrm{f-p}< H \leq H_\textrm{m},$\\
\frac{2-\gamma}{3-\gamma} \{ 2^{-\frac{1}{2-\gamma}}( h^{2-\gamma}+h_\textrm{m}^{2-\gamma})^{\frac{3-\gamma}{2-\gamma}}\\-(h_\textrm{m}^{2-\gamma}-1)^{\frac{3-\gamma}{2-\gamma}} -h^{3-\gamma} \}-h  & $,H_\textrm{a}<H \leq H_\textrm{m},$\\ 
\frac{2-\gamma}{3-\gamma} \{ ( h^{2-\gamma}+1)^{\frac{3-\gamma}{2-\gamma}} -h^{3-\gamma} \}-h & $,0 <H \leq H_\textrm{a},$ \\
\frac{2-\gamma}{3-\gamma} \left[ \{1-(-h)^{2-\gamma}\}^{\frac{3-\gamma}{2-\gamma}} -(-h)^{3-\gamma} \right]-h & $,-H_\textrm{f-p} < H \leq 0.$ }
\end{equation}
Here, $H_\textrm{m}$ is the maximum magnitude of external field where we change the sweep direction. To complete the hysteresis loop, $H_\textrm{m}$ is to be larger than $H_\textrm{f-p}$. $H_\textrm{f-p}$ is our fitting parameter, and above this value, the field profile keeps a `V'-shape (see left inset of Fig.~\ref{fig3}a), and as the field is reversed, it starts to be broken at certain points ($\pm x_b$) showing an upside-down `W'-shape. $H_\textrm{a}$ is the external field strength when the field profile finally becomes an upside-down `V'-shape (see right inset of Fig.~\ref{fig3}a). Since $x_b$ can be written in terms of $H$, $H_\textrm{f-p}$, and $H_\textrm{m}$, $H_\textrm{a}$ can be eliminated by the condition $x_b=d$. See Ref. 22 and Ref. 23 for more details. 

\ack
We are indebted to M. Nicklas, O. Stockert, J. D. Thompson, and S. Wirth for fruitful discussions. This work is supported by the Max Planck-POSTECH Center for Complex Phase Materials KR2011-0031558 and by the DFG Research Unit 960 Quantum Phase Transitions.


\section*{References}
\begin{thebibliography}{10}

\bibitem{Petrovic}
C. Petrovic, P. G. Pagliuso, M. F. Hundley, R. Movshovich, J. L. Sarrao, J. D. Thompson, Z. Fisk, and P. Monthoux, J. Phys.: Condens. Matter {\bf 13}, L337 (2001).

\bibitem{Bauer}
E. D. Bauer, C. Capan, F. Ronning, R. Movshovich, J. D. Thompson, J. L. Sarrao, Phys. Rev. Lett. {\bf 94}, 047001 (2005).

\bibitem{Paglione}
Johnpierre Paglione, M. A. Tanatar, D.G. Hawthorn, Etienne Boaknin, R.W. Hill, F. Ronning, M. Sutherland, Louis Taillefer, C. Petrovic, and P. C. Canfield, Phys. Rev. Lett. {\bf 91}, 246405 (2003).

\bibitem{Bianchi}
A. Bianchi, R. Movshovich, I. Vekhter, P. G. Pagliuso, and J. L. Sarrao, Phys. Rev. Lett. {\bf 91}, 257001 (2003).

\bibitem{Radovan}
H. A. Radovan, N. A. Fortune, T. P. Murphy, S. T. Hannahs, E. C. Palm, S. W. Tozer, D. Hall, Nature {\bf 425}, 51 (2003).

\bibitem{Movshovich}
R. Movshovich, M. Jaime, J. D. Thompson, C. Petrovic, Z. Fisk, P. G. Pagliuso, J. L. Sarrao, Phys. Rev. Lett. {\bf 86}, 5152 (2001).

\bibitem{Kenzelmann}
M.~Kenzelmann \textit{et al.,} Science {\bf 321}, 1652 (2008).

\bibitem{Young}
B. L. Young, R. R. Urbano, N. J. Curro, J. D. Thompson, J. L. Sarrao, A. B. Vorontsov, M. J. Graf, Phys. Rev. Lett. {\bf 98}, 036402 (2007).

\bibitem{Izawa}
K. Izawa, H. Yamaguchi, Yuji Matsuda, H. Shishido, R. Settai, Y. Onuki, Phys. Rev. Lett. {\bf 87}, 057002 (2001).

\bibitem{Daniel}
M. Daniel, E. D. Bauer, S. W. Han, C. H. Booth, A. L. Cornelius, P. G. Pagliuso, J. L. Sarrao, Phys. Rev. Lett. {\bf 95}, 016406 (2005).

\bibitem{Pham}
L. D. Pham, T. Park, S. Maquilon, J. D. Thompson, and Z.~Fisk, Phys. Rev. Lett. {\bf 97}, 056404 (2006).

\bibitem{Nair}
S.~Nair \textit{et al.,} Proc. Natl. Acad. Sci. {\bf 107}, 9537 (2010).

\bibitem{Booth}
C. H. Booth, E. D. Bauer, A. D. Bianchi, F. Ronning, J. D. Thompson, J. L. Sarrao, Jung Young Cho, Julia Y. Chan, C. Capan, and Z. Fisk, Phys. Rev. B {\bf 79}, 144519 (2009).

\bibitem{Urbano}
R. R. Urbano, B. L. Young, N. J. Curro, J. D. Thompson, L. D. Pham, Z. Fisk, Phys. Rev. Lett. {\bf 99}, 146402 (2007).

\bibitem{Bianchi1}
A. Bianchi, R. Movshovich, C. Capan, P. G. Pagliuso, and J.~L. Sarrao, Phys. Rev. Lett. {\bf 91}, 187004 (2003).

\bibitem{Tayama}
T. Tayama, A. Harita, T. Sakakibara, Y. Haga, H. Shishido, R. Settai, and Y. Onuki, Phys. Rev. B {\bf 65}, 180504(R) (2002).

\bibitem{Murphy}
T. P. Murphy, Donavan Hall, E. C. Palm, S. W. Tozer, C. Petrovic, Z. Fisk, R. G. Goodrich, P. G. Pagliuso, J. L. Sarrao, J. D. Thompson, Phys. Rev. B {\bf 65}, 100514(R) (2002).

\bibitem{Gratens}
X. Gratens, L. Mendonca-Ferreira, Y. Kopelevich, N. F. Oliveira Jr., R. R. Urbano, R. A. Ribeiro, R. Movshovich, J. L. Sarrao, J. D. Thompson, Z. Fisk, P. G. Pagliuso, Phys. Rev. B {\bf 85}, 054502 (2012).

\bibitem{Sakakibara}
T. Sakakibara, H. Mitamura, T. Tayama and H. Amitsuka, Jpn. J. App. Phys. {\bf 33}, 5067 (1994).

\bibitem{Larkin}
A. I. Larkin, Y. N. Ovchinnikov, J. Low Temp. Phys. {\bf 34}, 409 (1979).

\bibitem{Blatter}
G. Blatter, M. V. Feigel'man, V. B. Geshkenbein, A. I. Larkin, and V. M. Vinokur, Rev. Mod. Phys. {\bf 4}, 1125 (1994).

\bibitem{Bean}
C. P. Bean, Phys. Rev. Lett. {\bf 8}, 250 (1962).

\bibitem{Teruo}
T.~Matsushita, \textit{Flux Pinning in Superconductors} (Springer, 2007).

\bibitem{Bianchi2}
A. D. Bianchi \textit{et al.,} Science {\bf 319}, 177 (2008).

\bibitem{Mohan}
S. Mohan, J. Sinha, S.S. Banerjee, and Y. Myasoedov, Phys. Rev. Lett. {\bf 98}, 027003 (2007).

\bibitem{Paltiel}
Y.~Paltiel \textit{et al.,}
\newblock Nature {\bf 403}, 398 (2000).

\bibitem{Paltiel1}
Y. Paltiel, E. Zeldov, Y. Myasoedov, M. L. Rappaport, G. Jung, S. Bhattacharya, M. J. Higgins, Z. L. Xiao, E.Y. Andrei, P. L. Gammel, and D. J. Bishop, Phys. Rev. Lett. {\bf 85}, 3712 (2000).

\end{thebibliography}
\end{document}